\documentclass{optica-article}

\journal{opticajournal} % for journals or Optica Open

\articletype{Research Article}

\usepackage{lineno}
\usepackage{xcolor}
\usepackage{braket}
\usepackage{graphicx}

\begin{document}

\title{Decoy State-based Time Synchronization}

\author{Lukas Tiefenthaler\authormark{*}, Hannah Thiel, Davide Rusca, and Antia Lamas Linares}

\address{Vigo Quantum Communication Center, University of Vigo, Vigo E-36310, Spain; Escuela de Ingeniería de Telecomunicación, Department of Signal Theory and Communications, University of Vigo, Vigo E-36310, Spain; AtlanTTic Research Center, University of Vigo, Vigo E-36310, Spain\\
\authormark{*}ltiefenthaler@vqcc.uvigo.es}

\begin{abstract*} 
Time synchronization is a crucial requirement in quantum key distribution (QKD) protocols, ensuring accurate key generation via the correct assignment of bits of raw key and enabling eavesdropping detection via the precise recording of photon statistics. State-of-the-art experiments typically use an extra channel to synchronize the clocks of the transmitter and receiver via classical signals. In this work, we study the possibility of performing clock synchronization via the signals used for the key generation, which are already present in decoy-state-based BB84 protocols.

Without altering the protocol in any way, we use the different mean photon numbers of the signal and decoy states for time synchronization without a dedicated physical channel capable of clock synchronization. The proposed method relies only on the photons sent and received for key generation and does not require any change to the QKD protocol. The only change in the experiment is on the software level, thus making it very simple to implement. 

We demonstrate clock synchronization method in a simulation of a specific fiber-based QKD experiment. Like other decoy-state-based BB84 protocols, it is based on weak coherent pulses. In this simulation, we investigate the parameter space to find limits and optimal choices of our proposed method.

In addition to the non-protocol-altering clock synchronization method, we also discuss an approach that significantly improves performance in lossy channels by introducing an additional decoy state with a very high mean photon number.

By eliminating the need for an extra channel capable of clock synchronization, both methods proposed potentially reduce the complexity and cost of QKD systems and improve their agility.
\end{abstract*}

\section{Introduction}
Clock synchronization plays a central role in modern communication and distributed systems. Accurate timing coordination between remote nodes is essential for ensuring data integrity, minimizing latency, and enabling coherent operations across networks. In conventional systems, synchronization can often be achieved through classical techniques that exchange timing signals over dedicated channels. However, as communication technologies advance into domains such as QKD and quantum networks, traditional synchronization methods encounter fundamental limitations.

Quantum Key Distribution enables information-theoretical-secure key exchange using the principles of quantum mechanics \cite{RevModPhys.81.1301}. The BB84 protocol for coherent pulses with the addition of decoy states provides robust security against eavesdropping attacks \cite{Bennett_2014, Lo2005_decoy} and is currently the standard of most commercial QKD products. To correctly interpret photon arrival times and minimize errors, the clocks of the sender and receiver must be synchronized. In experiments with repetition rates that are in the region of 100$\,$MHz, the local oscillators can be synchronized using GNSS-disciplined local oscillators (GNSSDOs) \cite{BassoBasset2021}.

Since experiments are moving to higher repetition rates, the demands on the synchronization become too high to be satisfied by GNSSDOs. In many state-of-the-art experiments, the clocks are classically synchronized via a separate physical channel, which is not only expensive, but also prone to attacks \cite{calib_attack,man_in_the_middle,sync_attack}. Man-in-the-Middle attacks and an artificially introduced detector mismatch are possible when an eavesdropper can access the classical signal used for clock synchronization via the classical channel \cite{man_in_the_middle}. In QKD systems, a classical channel is always required, but it only needs to function as a logical channel, without requiring the immediate transmission of physical signals for clock synchronization. For QKD in fiber-poor environments, an extra channel capable of clock synchronization is hard to justify. In special cases like airborne and satellite-based QKD setups, such an extra channel comes with additional complexity and weight, which are often unwanted. 

The quantum channel of the original decoy-BB84 protocol provides enough information to achieve synchronization with no key production impact. Generally, all infrastructure needed in QKD setups that is not covered under the security proof poses a potential risk, so doing as much as possible within the bounds of the security proof is advantageous for a protocol's security.

Early clock synchronization methods in QKD using the quantum signals are  \cite{Marcikic2006, antia_sym_clock_sync, Scheidl_2009}, where the random patterns of entanglement-based QKD are exploited. For prepare and measure setups, the most widely discussed method for time synchronization via quantum Signals is the Qubit4Sync protocol \cite{PhysRevApplied.13.054041_Qbit4sync} and its variations \cite{Agnesi:20, fast_qubit-based_freqenzy-recovery}. A qubit-based synchronization method, especially for systems using time bin encoding, was demonstrated recently [\cite{first_timebin_qbit_based_sync}]. The practical implementation of such methods is demonstrated in real-time by using field-programmable gate arrays (FPGAs) in a protocol called IQSync  \cite{Krause2025}. Both, the Qubit4Sync and the IQSync methods rely on a published synchronization bit-string that is sent via the quantum channel and encoded in the more likely basis-choice of an asymmetrical decoy-BB84 protocol. 

Combining the two methods, \cite{Wang:21_timesync_satellite_decoy} demonstrated using the randomness of the vacuum (decoy) state to determine clock offset via the quantum channel without the need for a dedicated synchronization pattern. A broader approach is introduced by \cite{e23080988_bayesian}, where not only the randomness of the decoy state, but also the one of the basis choice is used for clock synchronization in a Bayesian approach. 

In contrast to those two methods, the protocol presented in this work is independent of the sifting process and relies only on the anyway published information about signal and decoy states. Additionally, we present an easy-to-implement method to greatly increase the clock synchronization performance in lossy  channels.

\section{Background}

\subsection{Decoy-State BB84 Protocol}
We tested the method on the example of a 1-decoy state BB84 protocol \cite{Bennett_2014, Lo2005_decoy, PhysRevA.72.012326} described in section \ref{sec:methodology}.
In the decoy-state method, Alice randomly varies the intensity of the coherent pulses she sends. In the one-decoy variation of the protocol, she chooses between high-intensity (signal) and low-intensity (decoy) pulses. After Alice reveals which pulses were decoys, Bob uses the corresponding measurement statistics to estimate the contribution of single-photon events and the information potentially gained by an eavesdropper, Eve. This estimation enables the application of privacy amplification to generate a secure key. Many modern QKD systems employ the decoy-state BB84 protocol \cite{davide_time_bin_encoding, Boaron2018}.

\subsection{Time Synchronization}
The synchronization of two remote clocks can be split into two parts.

\textbf{1. Frequency recovery}: The goal of this step is that the two local oscillators tick to the same frequency.  This syntonization can be achieved by locking on the arrival times of the cyclically sent signals. Depending on which local oscillator is used in the sender and receiver, the expected syntonization error can vary. While rubidium oscillators become more readily available, the local oscillators on most FPGAs are still quartz-based. For temperature-compensated quartz oscillators (TCXOs), typical values of 0.1 - 1$\,$ppm are reported.

\textbf{2. Time offset recovery}: Once the two local oscillators run at the same frequency, the clock offset can be determined and compensated. The time offset recovery can be performed cyclically to compensate for the drift of the two local oscillators when the syntonization is not perfect.

The time or clock offset recovery is considered to be the harder problem, and it is the focus of this work. 

\subsection{Time Synchronization Requirements in QKD}

Time synchronization in QKD ensures accurate interpretation of photon timestamps. Its performance is critical to reduce errors derived from misalignment \cite{RevModPhys.92.025002}. While a classical communications channel is necessary for all QKD experiments, the requirements for a channel capable of clock synchronization are far greater.

Depending on the repetition rate of the QKD experiment, the delay introduced by the synchronization channel must be stable to a certain degree. Networking cards, routers, and optical cross-connects present in ethernet-based channels introduce a variable lag in the \textmu s range. Long-term evolution (LTE)-based channels are only suited for timing accuracies of $1\,$ms in a best-case scenario. GNSS disciplined clocks can reach 10$\,$ns in a best-case scenario with reception at transmitter and receiver sites. Satellite-based methods like Two-Way Satellite Time Transfer are capable of achieving accuracies in the ns range. For modern QKD setups with repetition rates in the GHz, the period is $<1\,$ns, thus requiring more accurate time synchronization in the order of 10s of ps.

\subsection{Random Patterns for Clock Synchronization}
In clock synchronization, the clock offset can be determined from periodic synchronization signals, as is common in state-of-the-art experiments like the ones used for satellite missions \cite{Hiemstra2025_eagle1}.However, when the offset exceeds the signal period, this approach introduces ambiguity due to integer multiples of the period. Random patterns overcome this limitation: strong correlation with the received signal occurs only at the correct clock offset, making them uniquely effective for synchronization. Utilization of random patterns produced by photon pair sources has been reported for in \cite{Antia_clock_sync, lanning_time-sync, antia_sym_clock_sync, Scheidl_2009}. In prepare and measure setups, like the ones using a BB84 decoy protocol, states are usually prepared periodically. With only a small proportion of cases creating a transmitted photon, and no herald indicating when a photon was produced, these protocols are not naturally suited for clock offset recovery. However, by utilizing the random pattern superimposed on the periodic signal by the generation of decoy states, a unique clock offset can be obtained. In this work, we propose using the protocol-native random pattern for clock offset recovery.

\section{Methodology}
\label{sec:methodology}

\subsection{System Setup}

The main simulation models an existing QKD experiment  \cite{davide_time_bin_encoding, Boaron2018}, with the aim of finding a clock synchronization method that does not alter the protocol and utilizes only information that is already publicly available. In the modeled experiment, Bob reports back to Alice when a photon is detected with the additional information of how many time bins have passed since the last detection. Alice can thus reconstruct a detection pattern and can do the cross-correlation with the states she prepared. 

\begin{figure}[htbp]
\centering\includegraphics[width=7cm]{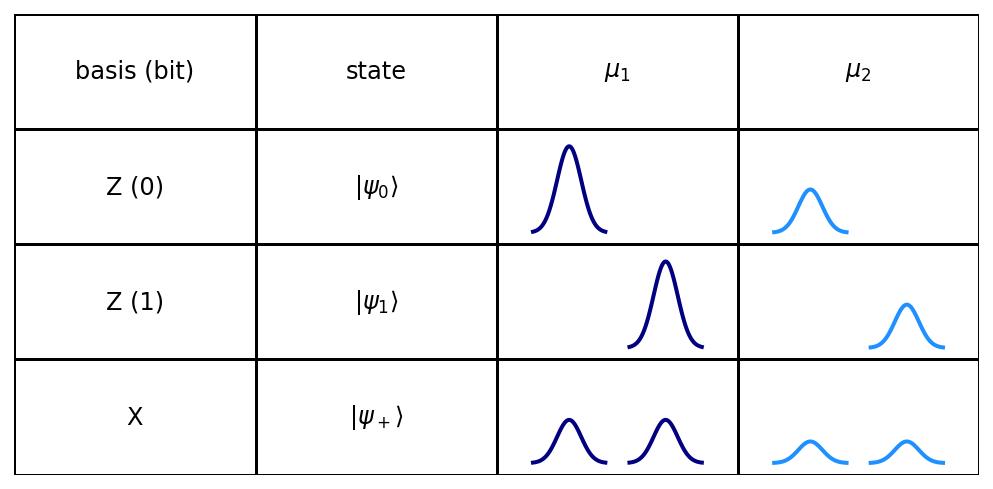}
\caption{Schematic of the encoding of states sent by Alice. Adapted from \cite{Boaron2018}.}
\label{fig:1-decoy}
\end{figure}

The protocol used in the experiment utilizes time bin encoding, where Alice can send one of three states, as shown in Fig. \ref{fig:1-decoy}.  In the Z basis, the states $ \ket{\psi_0} $ and $ \ket{\psi_1} $ are used to generate key, while the X basis with the single state $ \ket{\psi_+} $ is used to detect eavesdroppers.

\begin{figure}[htbp]
\centering\includegraphics[width=7cm]{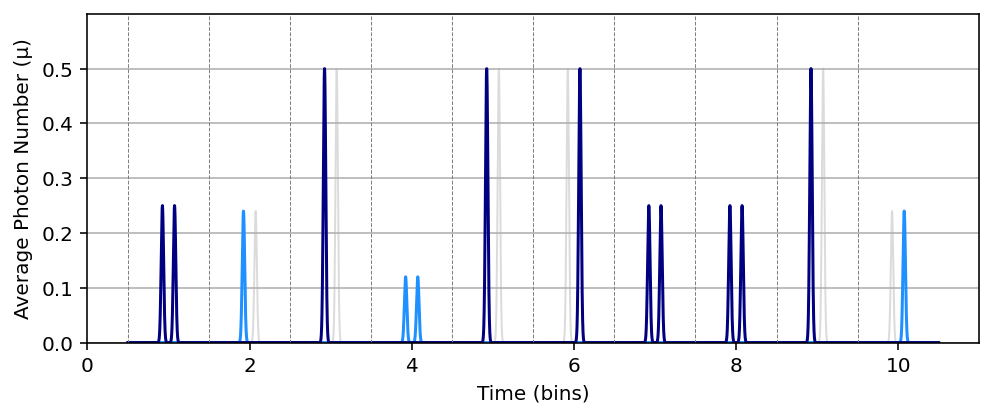}
\caption{\textbf{Decoy BB84 with time bin encoding}: A schematic of the pulses sent in a decoy BB84 protocol with time bin encoding. The different shades of blue show whether the coherent state was sent with high or low signal intensity (signal or decoy).}
\label{fig:BB84-pulses}
\end{figure}

A single decoy state is used, such that a qubit can be sent with one of two possible intensities \cite{davide_1-decoy}. A schematic of the random pattern of photon intensities sent via the quantum channel of the experiment can be seen in Fig. \ref{fig:BB84-pulses}. The double peaks hint at the fact that the experiment is using time bin encoding.

\subsection{Synchronization Technique}

The main part of the proposed time synchronization method is the cross-correlation of the states sent from Alice $a$ (signal or decoy) with the data received by Bob $b$. Those encoding-independent binary bit strings with different lengths are defined as:

a(i)=1(0) if Alice sends a signal (decoy) state

b(j)=1(0) if Bob receives (does not receive) a photon

The cross-correlation is given by 
\begin{equation}
(a \star b)(n) \;=\; \sum_{m=-\infty}^{\infty} \overline{a(m)}\, b(m+n),
\label{eq:crosscorr}
\end{equation}
where $n$ is indexing the time bins of the cross-correlation, and $m$ the ones of the initial signals. The result of the cross-correlation provides direct information about the offset between transmitter and receiver clocks, which can be obtained by finding the maximum correlation: 
\begin{equation}
\tau_{offset}=\text{argmax}((a\star b)(n))
\label{eq:argmax}
\end{equation}
To calculate the cross-correlation, a fast Fourier transform (FFT)-based algorithm was found to be the most efficient for the code running on a single core of a CPU without parallelization or other methods to optimize the calculation. In addition to the time offset, the syntonization error (frequency offset) of the clocks can be obtained by compressing and stretching one of the datasets while looking for the maximum correlation. 

\subsection{Simulated Data}
The simulation was performed to obtain data as close as possible to real experimental data while still having control of the parameters that influence the created data. The array of states  sent by Alice  $states_{\text{Alice}}$ is created by randomly choosing the predefined states (signal or decoy) with their corresponding probabilities:

$$S_1, S_2, \dots, S_{n_{\text{Alice}}} \sim \text{i.i.d. Categorical}(\{\mu_1, \mu_2\}, \{p(\mu_1), p(\mu_2)\})$$

$$\Pr(S_i = \mu_1) = p(\mu_1), \quad \Pr(S_i = \mu_2) = p(\mu_2), \quad i = 1, \dots, n_{\text{Alice}}$$
$$states_{\text{Alice}} = (S_1, S_2, \dots, S_{n_{\text{Alice}}})$$

where $\mu_1=0.5$, $\mu_2=0.25$,  $p(\mu_1)=0.7 $,  $p(\mu_2)=0.3 $ and $n_{\text{Alice}}$ is the block size chosen for the cross-correlation.  The random choice is performed via the function for independent and identically distributed random variables following a categorical distribution (i.i.d. Categorical()).

For the clock synchronization method, it does not make a difference in which detector (which basis) Bob detected a photon. So the simulation only needs a Boolean click-no-click array from Bob $Bob_{\text{click}}$, which is calculated by randomly choosing whether a click is detected or not using the following probability:

$$
P_{\text{click,n}} = 1 - e^{-{states_{\text{Alice, n}}} \cdot \eta} \cdot(1- bcr\cdot t_{\text{bin}} )
$$

where  $\eta$ corresponds to the channel transmissivity, $bcr$ to the background rate, and $t_{\text{bin}}$ to the length of a time bin. This click probability is calculated for every $\mu$ value in the extended array $states_{\text{Alice}}$. Note that $Bob_{\text{click}}$ is padded on both ends so that the cross-correlation with $states_{\text{Alice}}$ is computed only for positions where both arrays contain valid data.

\subsection{Suggested Real-Time Implementation on FPGAs}
\label{subsec:realtime_implementation}
Since many QKD experiments already use an FPGA for data handling, it would be beneficial to implement FFT-based cross-correlation on the already existing hardware. FFT-based cross-correlation efficiently computes the correlation in the frequency domain, reducing computational complexity from \(O(N^2)\) to \(O(N \log N)\), where $N$ is the signal length. 

The cross-correlation function shown in Eq. (\ref{eq:crosscorr}) is calculated using FFT via
$$(a \star b)(n) \;=\mathrm{IFFT}\{\,B[k]\cdot\overline{A[k]}\,\}[n]$$
where $A[k] = \mathrm{FFT}\{a[m]\}$ and $B[k] = \mathrm{FFT}\{b[m]\}$.

IFFT stands for the inverse fast Fourier transform. From the cross-correlation, the clock offset is recovered via Eq. (\ref{eq:argmax}).

FPGAs are well-suited for this implementation due to their ability to parallelize computations, resulting in low-latency and high-throughput performance.  Modern FPGA architectures also provide pre-optimized FFT cores that facilitate efficient hardware implementation \cite{Xilinx2020FFT, 1275737}. A real-time demonstration of clock synchronization via a predetermined synchronization pattern, where the calculations are performed on an FPGA, was presented in \cite{Krause2025}.

\section{Results and Analysis}
\label{sec:results}

For the fiber optical QKD experiment modeled in this work, the background rate is set to be 1$\,$kHz. According to \cite{davide_1-decoy}, this is a fair to pessimistic assumption for superconducting nano wire single-photon detectors (SNSPDs), whereas for avalanche photo diodes (APDs) it would be a relatively optimistic estimate. For a fiber-bound experiment, we assume a 100$\,$km long link with an attenuation of 0.2$\,$dB/km plus an additional 5$\,$dB of losses in the receiver, thus resulting in a total of 25$\,$dB of attenuation in the quantum channel. 

\begin{figure}[htbp]
\centering\includegraphics[width=7cm]{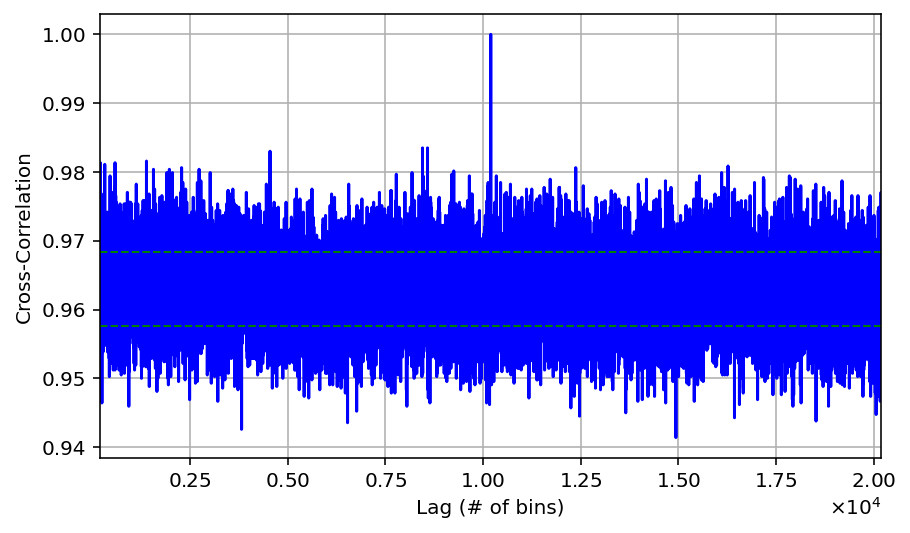}
\caption{\textbf{Normalized cross-correlation}: A representative cross-correlation at a channel loss of -25$\,$dB and a background rate of  $bcr = 1\cdot10^{4}$$\,$Hz. The green dashed lines show the standard deviation of the cross-correlation. The block size used was $n_{\text{Alice}} = 1\cdot 10^6$, and the quantum bit error rate (QBER) in this scenario was 21.1$\,$\%.}
\label{fig:cross_corr_fiber_sample}
\end{figure}

The block size of the prepared states that Alice uses for the cross-correlation is an important parameter of the synchronization method. Therefore, the success rate of the synchronization is investigated as a function of this block size while keeping other metrics of the QKD setup constant at reasonable values for a QKD setup using standard single-mode fiber and SNSPDs as detectors. Figure \ref{fig:cross_corr_fiber_sample} shows a representative cross correlation in which the delay was successfully found to be $1.1\cdot10^4$ bins. For this example, the frequency difference between the two local oscillators is considered to be 0. To ensure that a maximum is identified with probability greater than 99\% among $2\cdot10^4$ bins, the peak must lie approximately $5\sigma$ above the mean. In the example shown in Fig. \ref{fig:cross_corr_fiber_sample}, the maximum lies roughly $7\sigma$ above the mean.

\begin{figure}[htbp]
\centering\includegraphics[width=7cm]{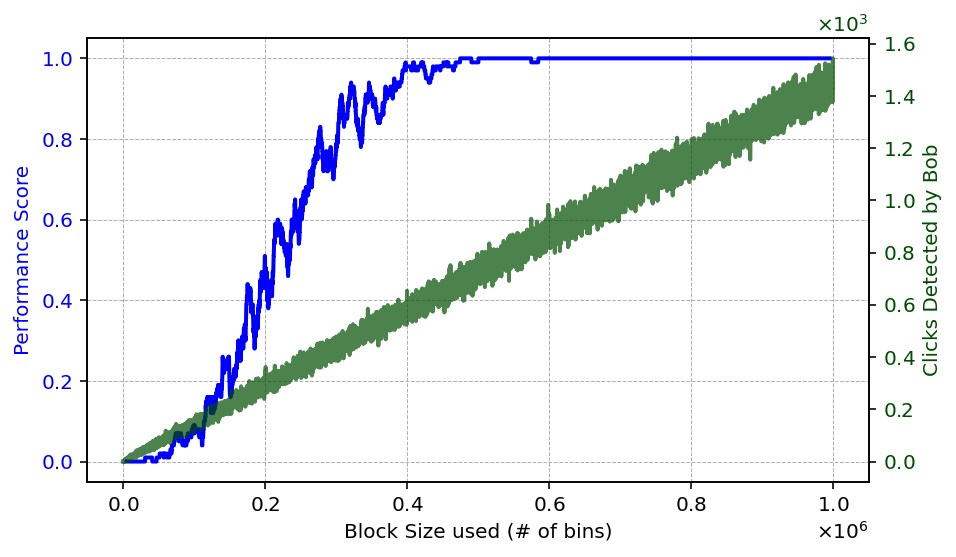}
\caption{\textbf{Cross-correlation performance as a function of block size}: Simulated clock synchronization performance at a channel loss of -25$\,$dB and a background rate of $bcr = 1$$\,$\,kHz.  The blue line shows the performance score of the clock synchronization, while the green curve shows the total detections by Bob.}
\label{fig:sync_succ_fiber_nAlice}
\end{figure}

\begin{figure}[htbp]
\centering\includegraphics[width=7cm]{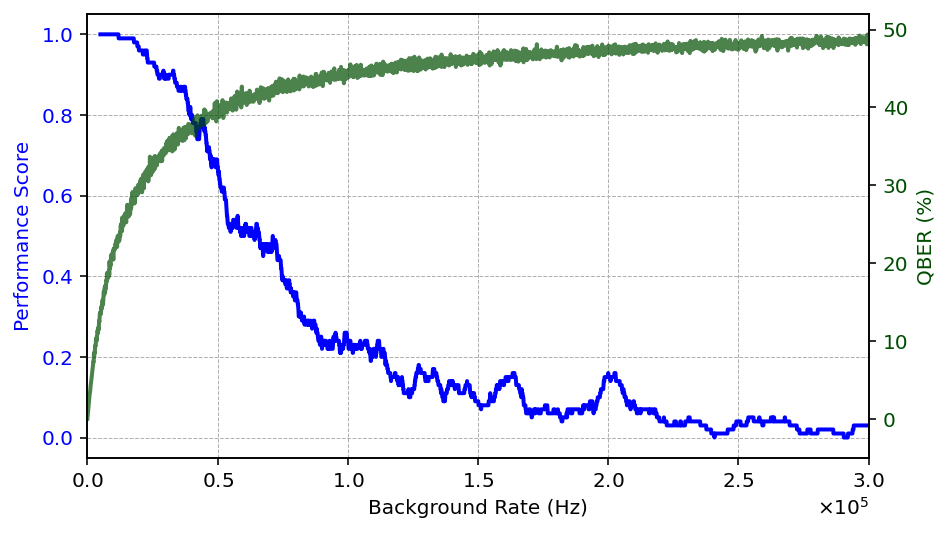}
\caption{\textbf{Cross-correlation performance as a function of background rate}: Clock synchronization performance as a function of the background rate. The simulation is carried out at a channel loss of -25$\,$dB and a block size of $n_{\text{Alice}} = 1\cdot 10^6$.  The blue line shows the performance score of the clock synchronization, while the green line represents the QBER.}
\label{fig:sync_succ_fiber_dcr}
\end{figure}

Figure \ref{fig:sync_succ_fiber_nAlice} shows the performance of the time synchronization as a function of the block size used. For the correct clock offset to be found successfully, Bob detects only a few hundred photons. With a channel loss of -25 dB and a backgroundrate of $bcr = 1$$\,$\,kHz, the regime in which the clock recovery works reliably starts at a block size of  $n_{\text{Alice}} = 5\cdot 10^5$. To benchmark the robustness of the method, the performance of the clock synchronization is plotted as a function of the background rate in Fig. \ref{fig:sync_succ_fiber_dcr}. 

Each simulation produces a binary result: 1 if the correct clock offset is identified and 0 otherwise. The performance score is obtained by averaging these binary outcomes over a centered sliding window of 100 simulations. As the window moves along the sequence of simulations, the score represents the local fraction of successful detections around each simulation index. While the performance score resembles a probability, it is an empirically smoothed success rate derived from observed simulation outcomes. Unlike a probabilistic outcome, it depends on the chosen window size and reflects local averaging rather than an underlying probability model. 

With an increasing background rate, the performance of the clock recovery decreases. From the plot in Fig. \ref{fig:sync_succ_fiber_nAlice}, we know that Bob detects $1.5\cdot 10^3$ legitimate states when $n_{\text{Alice}} = 1\cdot 10^6$.  When the background rate reaches $5\cdot10^4$$\,$Hz, the QBER reaches 40$\,$\%. At this QBER, the performance score is still 0.7. This shows how robust the proposed method is against the background and underlines its feasibility in noisy channels such as the ones present in satellite-based QKD. Notably, time synchronization can function even in regimes where secret key distribution is precluded. Interestingly, even when further increasing the background rate, the performance of the clock synchronization does not fully decrease to zero, but synchronization is still possible in a few percent of the cases.

\begin{figure}[htbp]
\centering\includegraphics[width=7cm]{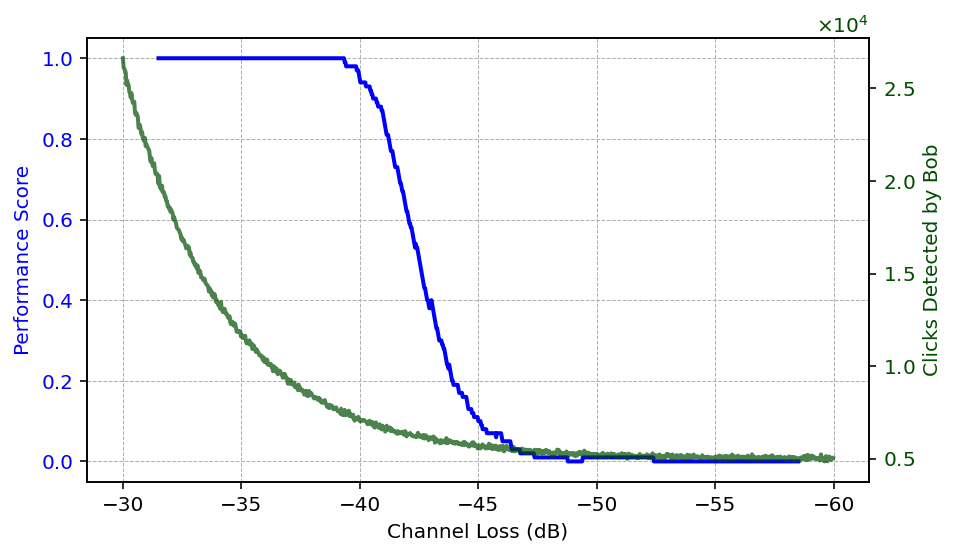}
\caption{\textbf{Cross-correlation performance as a function of channel loss}: Clock synchronization performance as a function of channel loss. The blue line shows the performance score of the clock synchronization, while the green line represents the number of detections from Bob. Alice's block size was $n_{\text{Alice}} = 5\cdot 10^7$, with a background rate of $dcr=1$$\,$kHz and a maximum clock offset of $\pm$3$\,$ms.}
\label{fig:sync_succ_fiber_channel_loss}
\end{figure}

Looking for the limit of the proposed method, we examine the highest channel loss at which this method can still function. We limit the block size to $n_{\text{Alice}} = 5\cdot 10^7$ as larger block sizes wouldn't be suitable for the IP cores of even very modern FPGA boards (see section \ref{subsec:feasablility_xilinx}). Figure \ref{fig:sync_succ_fiber_channel_loss} shows the resulting clock synchronization performance as a function of channel loss.

The results show that for realistic channel losses and background rates present in our fiber-based QKD experiment, the proposed method works very well for finding the clock offset. For a channel losses smaller then 40 dB the propesed method can work reliably with a performance score $> 90\,\%$. 
Generally high channel loss and background rates decrease the clock synchronization efficiency. A poor performance in clock synchronization can be improved by using larger block sizes for the cross-correlation, which increases the computational cost. In Fig. \ref{fig:sync_succ_fiber_nAlice}, we can observe that, at a reasonable background rate of $dcr=1$ kHz, the needed block size is comparable to the one needed for the Qubit4Sync method described in \cite{PhysRevApplied.13.054041_Qbit4sync}. While this means that the computational cost is similar, our proposed method is more efficient as the qubits used for synchronization do not have to be discarded. For a more detailed comparison between qubit based synchronization methods see Appendix1\_Benchmarking\_Table.

\section{Enhanced synchronization method for setups with high channel loss}
\label{sec:bright_pulses}

\begin{figure}[htbp]
\centering\includegraphics[width=7cm]{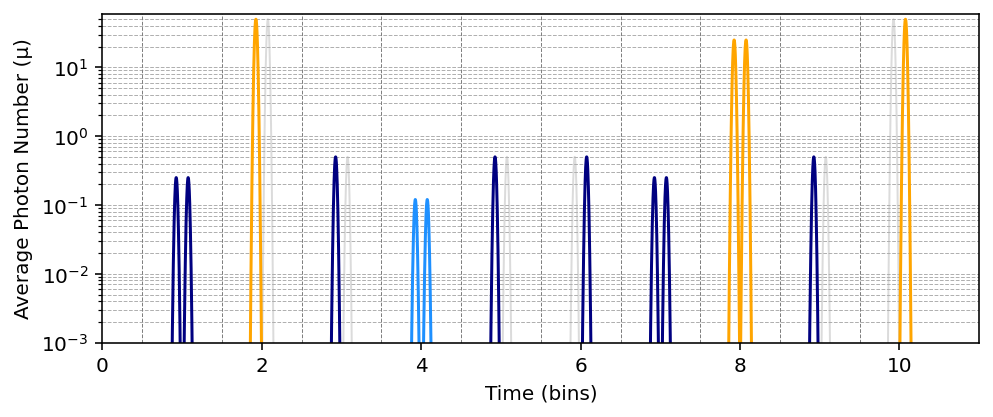}
\caption{\textbf{Decoy BB84 with synchronization pulses}: A schematic of the pulses sent in a decoy BB84 protocol with time bin encoding and bright synchronization pulses. The different shades of blue show whether the coherent state was sent with high or low intensity (signal or decoy). The orange  peaks represent the synchronization pulses sent with a higher intensity meant for clock synchronization.}
\label{fig:BB84-pulses-sync}
\end{figure}

In contrast to the previously proposed method, which leaves the protocol unchanged, we introduce a performance-boosting adaptation that involves only a minor modification to the protocol.  This alteration, which enhances clock synchronization performance in high-loss scenarios, is based on the use of an additional synchronization state that is bright yet transmitted with low probability.  The implementation of synchronization pulses effectively makes a 2-decoy BB84 protocol out of the original 1-decoy BB84 protocol, as shown in Fig. \ref{fig:BB84-pulses-sync}. 

The idea of using higher-intensity pulses for synchronization is similar to the protocol proposed by the EAGLE-1 mission \cite{Roessler_EAGLE1_TimeSync_2024}, with the key distinction that our method preserves their random pattern. In section \ref{sec:results_bright_pulses}, we demonstrate the method in our modeled experiment and compare it with the version that cross-correlates random patterns of decoy states without additional bright pulses.

\subsection{Results and Analysis of bright pulse method} 
\label{sec:results_bright_pulses}

\begin{figure}[htbp]
\centering\includegraphics[width=7cm]{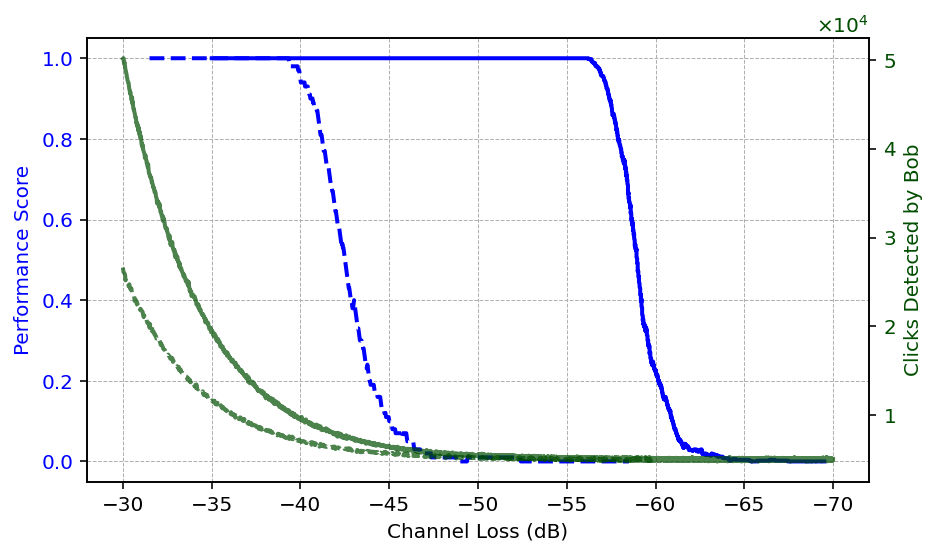}
\caption{\textbf{Cross-correlation performance as a function of channel loss}: This is a simulation of the impact of bright synchronization pulses on the clock synchronization. The channel loss is varied from -70$\,$dB to -30$\,$dB while $n_{\text{Alice}} = 5\cdot10^7$ and the background rate is 1$\,$kHz. The probability to send a synchronization pulse was $P(\mu_{\text{sync}})=0.01$, and the synchronization pulse brightness was $\mu_{\text{sync}} = 50$. The blue line shows the performance score of the clock synchronization, while the green line represents the number of detections from Bob. Data from Fig. \ref{fig:sync_succ_fiber_channel_loss} is displayed in dashed lines for comparison.}
\label{fig:sync_succ_pulse_channel_loss}
\end{figure}

The implementation of the additional synchronization pulses enables the clock synchronization at a given channel loss with smaller block sizes, thus making the method easier to implement in real-time setups. The effect of adding a synchronization pulse with a brightness of $\mu_{\text{sync}} = 50$ (20 dB above the signal state intensity $\mu_1=0.5$) and a send probability of $P(\text{sync})=0.01$ can be seen in Fig. \ref{fig:sync_succ_pulse_channel_loss}. All parameters were chosen to be the same as for the data shown in Fig. \ref{fig:sync_succ_fiber_channel_loss}, except for the adapted probabilities for signal, decoy, and synchronization states: $P'(\mu_1)=P(\mu_1)-P(\mu_{\text{sync}})/2$ and $P'(\mu_2)=P(\mu_2)-P(\mu_{\text{sync}})/2$
        
\begin{figure}[htbp]
\centering\includegraphics[width=7cm]{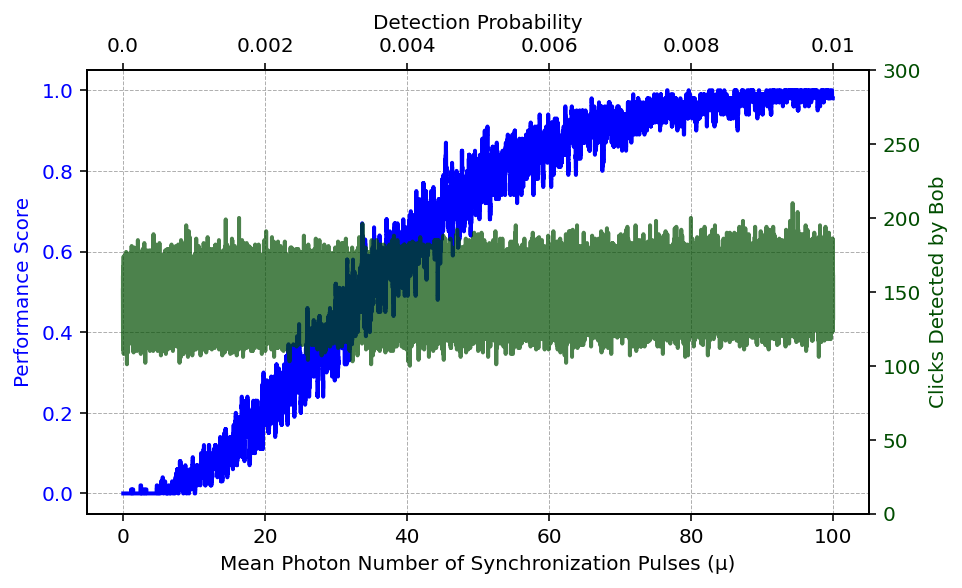}
\caption{\textbf{Cross-correlation performance as a function of synchronization pulse brightness}: This is a simulation of the impact of the brightness of the synchronization pulses on the clock synchronization via a channel with a loss of -40$\,$dB. The analyzed block size is $1\cdot10^6$ bins, and the background rate is 1$\,$kHz. The x-axis displays the mean photon number of the synchronization pulse when sent from Alice (lower scale) and the probability of Bob recording a synchronization pulse (upper scale). For the plot, the probability to send a synchronization pulse was set to be $P(\mu_{\text{sync}})=0.01$, while the synchronization pulse brightness was varied from $\mu_{\text{sync}} = 0$ to $\mu_{\text{sync}} = 100$. The blue line shows the performance score of the clock synchronization, while the green line represents the number of detections from Bob.}
\label{fig:sync_pulse_brightness}
\end{figure}

\begin{figure}[htbp]
\centering\includegraphics[width=7cm]{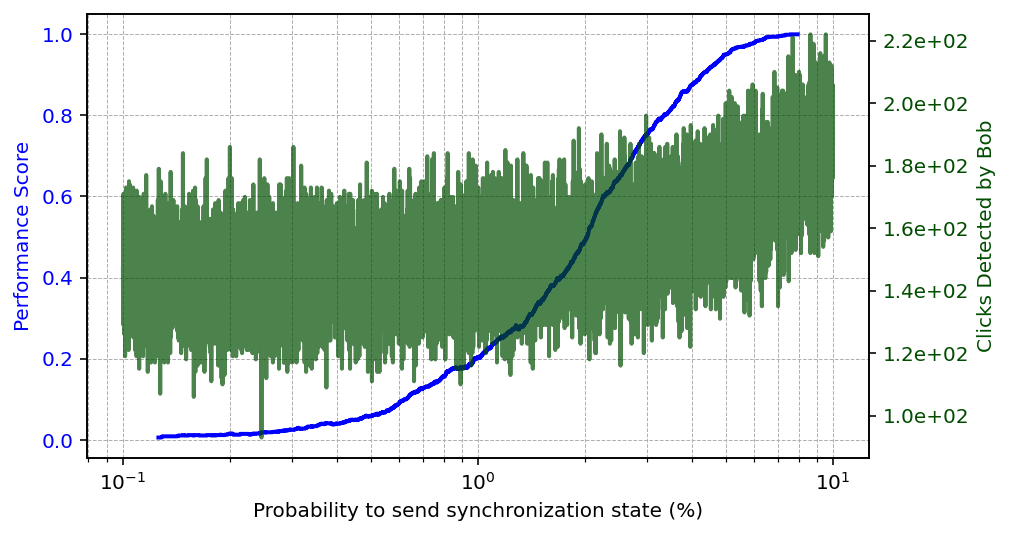}
\caption{\textbf{Cross-correlation performance as a function of synchronization pulse send probability}: This is a simulation of the impact of the brightness of the synchronization pulses on the clock synchronization via a channel with a loss of -40$\,$dB. The analyzed block size is $1\cdot10^6$ bins, and the background rate is 1$\,$kHz. The x-axis displays the probability $P(\mu_{\text{sync}})$ with which a synchronization pulse is sent. It is varied from $P(\mu_{\text{sync}})=0.1$\% to $P(\mu_{\text{sync}})=10$\%. The signal intensity of the synchronization pulse was set to be $\mu_{\text{sync}} = 5$.}
\label{fig:sync_pulse_send_probability}
\end{figure}

In Fig. \ref{fig:sync_pulse_brightness}, clock synchronization is simulated using synchronization pulses with the probability of being sent $P(\mu_{\text{sync}})=0.01$. The performance of the clock offset recovery is shown as a function of synchronization pulse brightness $\mu_{\text{sync}}$. Similarly, the synchronization can be improved by sending synchronization pulses with a higher repetition rate but lower intensity, as shown in Fig. \ref{fig:sync_pulse_send_probability}. Here, the brightness of the synchronization pulse was fixed to be $\mu_{\text{sync}}=5$, which is 10 dB higher than the signal intensity $\mu_1=0.5$.  

The bright pulse method increases the window in which clock synchronization is possible by roughly 17$\,$dB of channel loss, allowing us to perform clock offset recovery with a performance score of  0.5 at a channel loss of -59$\,$dB instead of  -42.5$\,$dB.

In scenarios where the channel loss is lower, adopting the bright pulse method allows for a reduction of the block sizes used for the clock offset recovery. This way, the block sizes reach sizes similar to the bit string length needed in the distributed frame synchronization method proposed in \cite{chen2024qubitbaseddistributedframesynchronization} that is based on the Qubit4Sync method \cite{PhysRevApplied.13.054041_Qbit4sync}. The computational cost should therefore also be comparable, as the operations are very similar. Using the bright pulse method, however, the bins used for the clock offset recovery can still be used for key generation. The penalty on the key rate is only dependent on $P(\P_{\text{sync}})$ as discussed in section \ref{subsec: cost of bright pulses} (Appendix1\_Benchmarking\_Table).

\section{Discussion}
\label{sec:discussion}

\subsection{Computational limitations - maximum clock offset}
Since the simulation is currently carried out on a desktop PC in a non-real-time application, there is almost no limit to the maximum clock offset $d_{\text{max}}$. Whether the correct offset is found in the cross correlation is more dependent on channel loss, block size, and background rate. It plays a major role, however, in the speed of the calculation of the cross-correlation. So, for the implementation on an FPGA for a real-time application, the factor should be as small as possible.

The maximal clock offset is only large at the initial clock offset analysis. In all subsequent calculations, the maximum offset between the clocks $d_{\text{max}}$ is bounded by the maximum drift of the two local oscillators, which can be monitored and corrected in real time with a small computational overhead.

\subsection{Computational limitations - maximum block size}
\label{subsec:comp_lim}
In the experiment described in section \ref{sec:methodology}, each detected photon with its corresponding relative time difference since the last click is reported to Alice via the classical channel. When the QKD setup is generating key, Alice uses this information to immediately delete all her stored information about the qubits sent in the bins where Bob detected no photon. For this to work, however, the clocks of Alice and Bob must already be synchronized. So, for the initial synchronization, no data can be deleted, thus limiting the maximum size of the signal block that Alice can compare with the signal reported from Bob, to what is possible in terms of computing power and memory of the FPGA board used.  The channel delays also play a crucial role in this regard, since the information about which signals were sent needs to be stored until the detected signal is reported back by Bob. The resulting storage time of a block of sent states is thus the sum of the latencies of the classical and quantum channels.

The performance of the specialized FFT IP cores present on most modern FPGA boards decreases as the FFT array size increases. The two factors influencing the array size are the maximum delay between the clocks $d_{\text{max}}$ , as well as the needed block size $n_{\text{Alice}}$. The maximum clock offset is given by the local oscillators that are available, so the parameter that can be optimized to enhance the computing performance, and with it the real-time feasibility of the proposed method, is $n_{\text{Alice}}$. The block size of the sent states Alice uses to perform the cross-correlation with the signal reported back by Bob is therefore one of the crucial parameters of this synchronization process. Its ideal value varies depending on parameters of the quantum channel, such as channel loss, background rate, and also the syntonization error of the two local oscillators.

\subsection{Computational feasibility with existing FPGA devices}
\label{subsec:feasablility_xilinx}
The Xilinx Kintex-7 and UltraScale FFT IP cores support a maximum transform length of $2^{27}$ points \cite{Xilinx2020FFT}. For a repetition rate of $2.5\,\text{GHz}$, the inequality
\begin{equation}
    n_{\text{Alice}} + 2\cdot d_{\max} \leq  2^{27}  
\end{equation}
must be met. For the maximal block size $n_{\text{Alice}} = 5\cdot10^7$ used in this work, the hardware's maximal tolerable clock offset would be $\pm16.8\,\text{ms}$ or $d_{\max} = 4.2\cdot10^7$ bins.\\
For block sizes of $n_{\text{Alice}} = 5\cdot10^5$, sufficient for channel losses below $25\,\text{dB}$ (see figure \ref{fig:sync_succ_fiber_nAlice}), the required transform length reduces to $N = 5\cdot10^5 + 2\cdot d_{\max}$. For $d_{\max} = \pm3\,\text{ms}$ this gives $N \approx 1.5\times10^7 < 2^{24}$, which would be achievable with lower tier FPGAs.\\
Additional constraints on the maximum feasible block size arise from the available BRAM capacity on the FPGA and the 
classical channel latency, which determines the minimum buffer size for storing Alice's sent states while awaiting Bob's detection reports. A precise estimation of these limits is hardware- and setup-specific and is left for future work.

\subsection{Robustness to frequency offsets}
In realistic QKD systems, residual frequency offsets between Alice's and Bob's local oscillators are unavoidable. For TCXOs with a stability of 0.1-1\,ppm, a frequency offset of 1\,ppm over a block of $n_{\text{Alice}} = 10^6$ bins at 2.5\,GHz corresponds to a shift of 1 bin across the block, smearing the cross-correlation peak. For $n_{\text{Alice}} = 5\cdot10^7$ this increases to 50 bins. A natural mitigation is to lock Bobs local oscillator on the arrival times of detected photons, which is effective down to approximately - 44\,dB of channel loss ($\sim 10$ detections per bin shift). Below, the photon stream becomes too sparse for reliable drift estimation. This aligns well with the performance limits of the cross-correlation displayed in figure \ref{fig:sync_succ_fiber_channel_loss}.
In the high-loss regime above this limit, the bright synchronization pulses introduced in Section~\ref{sec:bright_pulses} provide a natural solution: since they arrive at Bob with a detection probability orders of magnitude higher than signal or decoy states, they constitute a much denser and more reliable timing reference for arrival-time locking, effectively extending the range of viable syntonization to the same channel loss regime in which the bright pulse method enables clock offset recovery.

\subsection{Time bin filtering during synchronization}
Defining a meaningful background rate for free-space QKD setups is challenging, as it depends, apart from the telescope used, on parameters like the exact location on Earth, day time, lunar phase, seeing angle, turbulence, etc. A table of measured rates using a defined set of filters was published by Er-long et. al \cite{Er-long_2005_free_space_background}. 

Proposed methods to reduce the background rate in free-space experiments, such as time bin filtering during daytime QKD, can greatly reduce the background rate. The filtering window can, however, only be narrowed down after the clocks are fully synchronized, thus providing less effective filtering during the clock offset recovery. For reducing the background rate during the synchronization process, we have to rely on timing-independent procedures such as narrow band pass filters and adaptive optics in combination with a narrow field stop \cite{lanning_sky_radiance_day}. 

\subsection{Impact of bright synchronization pulses on keyrate}   
\label{subsec: cost of bright pulses}

In principle, the bright pulses could be included in the security proof and treated as just another bright decoy state. The penalty resulting from this through the privacy amplification is, however, higher than if the pulses are directly disregarded.  In a scenario where all synchronization states are excluded from the key generation, and the probability of sending a synchronization state $P(\mu_{\text{sync}})=0.01$, the penalty on key generation is also 1\%.

In the modeled experiment, however, the dead time of the detector is longer than the period of the signal, so the detector is blinded for the following qubits. Our signal, with a repetition rate of 2.5$\,$GHz, has a period of 400$\,$ps. Assuming a deadtime of 10$\,$ns, this means that the detector is blind for the following 25 qubits. In Fig. \ref{fig:sync_pulse_brightness}, the probability of the detection of a synchronization pulse is shown for a total channel loss of -40$\,$dB. For the highest intensity of the synchronization pulses of $\mu_{\text{sync}} = 100$, the detection probability is $P(detect\, sync)=0.01$. So the total sacrifice of key in that case would be 1.25$\,$\%.
The exclusion of synchronization bins from key generation is justified on the following grounds. First, synchronization states carry no encoded key bit or basis information, so their interception yields Eve no advantage in guessing the key. Second, Alice publicly announces the positions of synchronization bins, as she does for decoy state labels. So Eve's knowledge of these positions is already assumed in the security analysis. Third, since the synchronization bins are selected randomly and independently by Alice, Eve cannot exploit their positions to mount a selective attack on neighboring key-generating bins beyond what is already bounded by the dead time analysis above. Eve could therefore at most use the bright synchronization pulses to disrupt the synchronization of Alice’s and Bob’s clocks. This capability is equivalent to attacks on the classical channel and would only manifest as an increased QBER. 
The cost of our proposed method in terms of secret key rate can be understood as demonstrated in Table \ref{tab: cost of bright pulses}.

\begin{table}[htbp]
\centering
\caption{A table summarizing the cost of our proposed synchronization methods in high channel loss scenarios. At some point, the needed block size gets too large to be handled by an FPGA in real-time. Hence, it is necessary to change to the method using bright synchronization pulses. While those bright synchronization pulses have an impact on the key rate it can be minimized by increasing the brightness of the synchronization pulses. When the experimental setup does not allow for very bright pulses, the send probability can be increased to cope with higher loss, but at an increased penalty on the key rate.}
\label{tab: cost of bright pulses}
\begin{tabular}{|c|c|c|c|}
\hline
\textbf{method} & \textbf{key rate cost} & \textbf{$\uparrow$ channel loss} & \textbf{setup cost} \\ \hline
no bright pulses & 0  & increase $n_{\text{Alice}}$ & cpu power \\ \hline
bright pulses &  $\uparrow$ $\rho$ \% & $\uparrow$ $P(\mu_{\text{sync}})$ &0\\ \hline
brighter pulses & $\downarrow$ $\rho$ \% & $\uparrow$ $\mu$ & attenuator range\\ \hline
\end{tabular}
\end{table}

\subsection{Advantages of our proposed bright pulse method}
One of the benefits of this method over classical periodic reference pulses is, that the advantage of the unequivocal fit of the random pattern cross-correlation remains. Since the bright synchronization pulses are still very weak in terms of classical signal brightness, they can still be realized with the intensity modulator and variable attenuators present in typical prepare and measure setups. For the practical implementation of cross-correlation on an FPGA, one could also use smaller block sizes, enabling a more computationally efficient setup.

\subsection{Compatibility of bright synchronization pulses with other QKD protocols}
A photon number of $\mu_{\text{sync}}=50$ with a send probability of $P(\text{sync})=0.01$ leads to the reduction of the block size by roughly two magnitudes. A synchronization pulse of this brightness is still achievable with intensity modulators, which are available in prepare and measure setups without the need for extra hardware. Depending on the requirements of the setup,  a lower brightness $\mu_{\text{sync}}$ can be used by increasing  $P(\text{sync})$ and vice versa. 

Cross-correlation is possible with many different QKD setups, but bright synchronization pulses may be difficult in setups using continuous wave, as the intensity is usually not a parameter controlled by the experimenter. In pulse-based entanglement QKD systems, the method could be implemented by multi-pair generation.

\subsection{Bright pulse method for low earth orbit (LEO) satellite approach}
In some QKD setups, the channel loss and background rates are variable. Our proposed cross-correlation-based synchronization scheme works even in setups with a QBER excessive for effective key generation. In some cases, it may still be desirable for the clock synchronization to work even while channel loss is extremely high. One such case is the approach of a LEO QKD satellite. One could use the approach in which the altitude angle and the transmissivity are low to establish a stable synchronization of the transmitter and receiver clocks before the start of the key generation at higher altitude angles. This maximizes both the time available for key generation and the transmitted key length in one path, allowing for more efficient error correction and privacy amplification. 

\section{Conclusion}
This paper demonstrates how the existing hardware and functionality of the quantum channel in a standard decoy BB84 implementation can be utilized to achieve synchronization without requiring any additional equipment. Because the protocol itself remains unmodified, our synchronization method fully preserves compliance with the established security proofs. For fiber-based setups with channel losses smaller than 30$\,$dB, the proposed method achieves high reliability with no impact on QKD performance, since it uses existing data from the protocol working in its standard form. For higher noise and channel loss experiments, such as those featuring free space links, the addition of a bright synchronization state to the BB84 decoy protocol seems highly beneficial for the clock synchronization while still having only a minor impact on the QKD performance.

\begin{backmatter}
\bmsection{Funding}
This work was supported by the Galician Regional Government (consolidation of research units: atlanTTic), the Spanish Ministry of Science, Innovation and Universities through the grant No. PID2024-162270OB-I00, the “Hub Nacional de Excelencia en Comunicaciones Cuanticas” funded by the Spanish Ministry for Digital Transformation and the Public Service and the European Union NextGenerationEU, the European Union’s Horizon Europe Framework Programme under the project “Quantum Secure Networks Partnership” (QSNP, grant agreement No 101114043), the European Union via the European Health and Digital Executive Agency (HADEA) under the Project QuTechSpace (grant 101135225), the European Union under the project Project IberianQCI (grant 101249593), as well as the Programa de Cooperación Interreg VI-A España–Portugal (POCTEP) 2021–2027 through the project QUANTUM IBER$\_$IA.\\
This research was funded in whole or in part by the Austrian Science Fund (FWF) 10.55776/J4947. For open access purposes, the author has applied a CC BY public copyright license to any author-accepted manuscript version arising from this submission.

\bmsection{Disclosures}
The authors declare no conflicts of interest.

\bmsection{Data Availability Statement}
The code used to perform the simulations is available on Github: \url{https://github.com/VQCC-Satellite/TimeSync_decoy-BB84}. The datasets presented in this paper are not publicly available at this time but may be obtained from the authors upon reasonable request.\\
\\
See Supplement 1 for supporting content.

\bibliography{mybib}
\end{backmatter}

\newpage
\appendix
\section{Supplement 1: Benchmarking table for prepare and measure - qubit based clock synchronization methods}

Comparison of clock synchronization methods for prepare-and-measure 
QKD. Block sizes are given for the channel loss conditions reported. The computational 
complexity of all FFT-based methods scales as $\mathcal{O}(N \log N)$ with the block 
size $N$. The IQSync method uses a fundamentally different approach with a polylogarithmic 
average-case recovery complexity of $\mathcal{O}((\log_2 \Delta_{\max})^b)$, where 
$\Delta_{\max}$ is the maximum recoverable offset and $b \approx 1.1$ ($b \approx 3.0$) 
for no (maximum) interleaving~\cite{Krause2025}; the block size quoted for IQSync 
therefore reflects the total pattern duration rather than a cross-correlation window.
$^\dagger$Qubits used for synchronization are discarded from key generation.
$^\ddagger$Protocol requires minor modification: an additional bright decoy state is 
introduced, effectively extending the standard 1-decoy BB84 to a 2-decoy scheme.
$^\S$Channel loss estimated from reported count rates of 1000-28000\,cps at a 
repetition rate of 100\,MHz, see text.
$^\P$Block size estimated from the reported accumulation time of 0.2\,s at 
100\,MHz repetition rate.
$^\#$Block size of 33,110 sampling bin widths (corresponding to 4,140 communication 
bin widths at $m = 8$ sampling bins per communication bin) achieves 95\% 
synchronization confidence at $\mu = 0.01$ and $d = 8\times10^{-4}$~\cite{e23080988_bayesian}.
$^\star$Key rate penalty depends on the ratio of synchronization pattern length 
to total session length and the resynchronization frequency; 
values quoted assume a single synchronization per key generation session.
$^\circ$Results based on simulation only.
\begin{table}[ht]
\centering
\begin{tabular}{lcccccc}
\hline
\textbf{Method} & \textbf{Channel loss} & \textbf{Block size} & \textbf{Qubits} & 
\textbf{Protocol} & \textbf{Rate} & \textbf{Key rate} \\
 & \textbf{(dB)} & \textbf{(bins)} & \textbf{discarded} & \textbf{modified} & 
\textbf{(MHz)} & \textbf{penalty (\%)} \\
\hline
Qubit4Sync~\cite{PhysRevApplied.13.054041_Qbit4sync}   & 19.0  & $10^6$            & Yes$^\dagger$ & Yes   & 50    & $\sim$0-5$^\star$          \\
Qubit4Sync~\cite{PhysRevApplied.13.054041_Qbit4sync}   & 38.0  & $10^7$            & Yes$^\dagger$ & Yes   & 50    & $\sim$0-5$^\star$          \\
IQSync~\cite{Krause2025}                               & 71.2  & $1.56\cdot10^{10}$& Yes$^\dagger$ & Yes   & 625   & session-dep.$^\star$        \\
Dist.\ Frame~\cite{chen2024qubitbaseddistributedframesynchronization}  & 29.7  & $10^5$            & Yes$^\dagger$ & Yes   & 50    & session-dep.$^\star$   \\
Dist.\ Frame~\cite{chen2024qubitbaseddistributedframesynchronization}  & 29.2  & $10^4$            & Yes$^\dagger$ & Yes   & 625   & session-dep.$^\star$   \\
Vacuum state~\cite{Wang:21_timesync_satellite_decoy}   & 31-46$^\S$ & $2\cdot10^7$$^\P$ & No       & No    & 100   & 0         \\
Bayesian~\cite{e23080988_bayesian}$^\circ$             & 13    & $3.3\cdot10^4$$^\#$& No            & No    &       & 0         \\
\hline
This work (no bright pulses)$^\circ$                   & 25    & $5\cdot10^5$      & No            & No    & 2500  & 0         \\
This work (no bright pulses)$^\circ$                   & 40    & $5\cdot10^7$      & No            & No    & 2500  & 0         \\
This work (bright pulses)$^\circ$                      & 57    & $5\cdot10^7$      & Yes$^\dagger$  & Yes$^\ddagger$ & 2500 & $\sim$1.25 \\
\hline
\end{tabular}
\end{table}

\end{document}